\documentclass[amsmath,amssymb,floatfix,lengthcheck,nopreprintnumbers,prl,showpacs,superscriptaddress,twocolumn]{revtex4}
\usepackage{graphicx}

\newcommand{\beq}{\begin{equation}}
\newcommand{\eeq}{\end{equation}}
\newcommand{\beqs}{\begin{eqnarray}}
\newcommand{\eeqs}{\end{eqnarray}}

\newcommand{\pbp}[0]{\ensuremath{\langle \overline{\psi} \psi \rangle}}

\newcommand{\tr}{{\rm tr}}
\def\CPT{{$\chi$PT$\,$}}
\def\EWC{{EW$\chi$L$\,$}}

\begin{document}

\title{WW Scattering Parameters via Pseudoscalar Phase Shifts}

\author{T.~Appelquist}
\affiliation{Department of Physics, Sloane Laboratory, Yale University,
             New Haven, Connecticut 06520, USA}
\author{R.~Babich}
\affiliation{Department of Physics, Boston University,
	Boston, Massachusetts 02215, USA}
\author{R.~C.~Brower}
\affiliation{Department of Physics, Boston University,
	Boston, Massachusetts 02215, USA}
\author{M.~I.~Buchoff}
\affiliation{Physical Sciences Directorate, Lawrence Livermore National Laboratory,
	Livermore, California 94550, USA}
\author{M.~Cheng}
\affiliation{Physical Sciences Directorate, Lawrence Livermore National Laboratory,
	Livermore, California 94550, USA}
\author{M.~A.~Clark}
\affiliation{Harvard-Smithsonian Center for Astrophysics, Cambridge, Massachusetts 02138, USA}
\author{S.~D.~Cohen}
\affiliation{Department of Physics, University of Washington, Box 351560,
               Seattle, WA 98195, USA}
\author{G.~T.~Fleming}
\affiliation{Department of Physics, Sloane Laboratory, Yale University,
             New Haven, Connecticut 06520, USA}
\author{J.~Kiskis}
\affiliation{Department of Physics, University of California,
	Davis, California 95616, USA}
\author{M.~F.~Lin}
\affiliation{Department of Physics, Sloane Laboratory, Yale University,
             New Haven, Connecticut 06520, USA}
\author{E.~T.~Neil}
\affiliation{Theoretical Physics Department, Fermi National Accelerator Laboratory,
             Batavia, IL 60510, USA}
\author{J.~C.~Osborn}
\affiliation{Argonne Leadership Computing Facility,
	Argonne, Illinois 60439, USA}
\author{C.~Rebbi}
\affiliation{Department of Physics, Boston University,
	Boston, Massachusetts 02215, USA}
\author{D.~Schaich}
\affiliation{Department of Physics, Boston University,
	Boston, Massachusetts 02215, USA}
\affiliation{Department of Physics,
University of Colorado, Boulder, CO 80309, USA}
\author{S.~Syritsyn}
\affiliation{%
Lawrence Berkeley National Laboratory, Berkeley, CA 94720, USA}
\author{G.~Voronov}
\affiliation{Department of Physics, Sloane Laboratory, Yale University,
             New Haven, Connecticut 06520, USA}
\author{P.~Vranas}
\affiliation{Physical Sciences Directorate, Lawrence Livermore National Laboratory,
	Livermore, California 94550, USA}
\author{J.~Wasem}
\affiliation{Physical Sciences Directorate, Lawrence Livermore National Laboratory,
	Livermore, California 94550, USA}
\collaboration{Lattice Strong Dynamics (LSD) Collaboration}
\noaffiliation

\preprint{LLNL-JRNL-499587}

\begin{abstract}
Using domain-wall lattice simulations, we study pseudoscalar-pseudoscalar scattering in the maximal isospin channel for an $SU(3)$ gauge theory with two and six fermion flavors in the fundamental representation. This calculation of the S-wave scattering length is related to the next-to-leading order corrections to WW scattering through the low-energy coefficients of the chiral Lagrangian. While two and six flavor scattering lengths are similar for a fixed ratio of the pseudoscalar mass to its decay constant, six-flavor scattering shows a somewhat less repulsive next-to-leading order interaction than its two-flavor counterpart.   Estimates are made for the WW scattering parameters and the plausibility of detection is discussed.
\end{abstract}


\maketitle
%

\section{Introduction}
Technicolor theories based on scaled-up QCD may be disfavored due to large contributions to the electroweak S-parameter and a chiral condensate too small to account for flavor effects \cite{Eichten:1979ah}. However, walking \cite{Holdom:1981rm,Bando:1986bg,Appelquist:1986an} could account for these differences.   For a gauge theory with a fixed number of colors, walking is believed to set in at a number of flavors $N_f$ just below the minimum number $N_f^c$ required to generate an infrared fixed point.  Above $N_f^c$, the theory is conformal in the infrared, and this behavior persists until asymptotic freedom is lost.

Since walking and the onset of infrared conformality involve strong dynamics, numerical lattice field theory is ideally suited to study this behavior. To this end, lattice calculations \cite{Catterall:2009sb,Hasenfratz:2009ea,DelDebbio:2010hx,Fodor:2011tu,DelDebbio:2011rc} have been performed with different numbers of flavors, colors, and in different representations, leading to evidence for both confining and conformal behavior, with much debate as to the boundaries of these phases.  The present work builds on two earlier papers of the LSD collaboration \cite{Appelquist:2009ka,Appelquist:2010xv}, where evidence for both chiral condensate enhancement and dynamic reduction of the S-parameter were observed for an $SU(3)$ gauge theory when $N_f$ is increased from $2$ to $6$, still well below $N_f^c$.

In this work, using domain-wall fermions, we make the first lattice connection to WW scattering for these two theories. This process, whose longitudinal modes can be related to scattering of pseudo-Nambu-Goldstone bosons (PNGBs) via the equivalence theorem,  is receiving renewed attention.  As a direct probe of the physics behind electroweak symmetry breaking, WW scattering could be an important channel to investigate at the LHC.

Our approach is to compute the parameters of effective, low-energy chiral Lagrangians. We focus on two coefficients of the electroweak chiral Lagrangian \cite{Longhitano:1980iz,Appelquist:1993ka} which encode the dominant deviations of
the longitudinal WW scattering amplitude from that of the standard model \cite{Bagger:1995mk,Distler:2006if,Vecchi:2007na}. These coefficients, in turn, are related to certain coefficients of the $N_{f}$-flavor hadronic chiral Lagrangian \cite{Gasser:1983yg,Gasser:1984gg}. For QCD ($N_f = 2$),
there are two such parameters ($l_1$ and $l_{2}$). Lattice methods have so far constrained only linear combinations of these terms via pion form factors \cite{Aoki:2009qn} or the extraction of effective range parameters in I=2 $\pi\pi$ scattering \cite{Beane:2011sc}.

Here we calculate the leading term (scattering length) in the effective-range expansion  for S-wave, maximal-isospin pseudoscalar scattering for the $N_f = 2$ and $N_f = 6$ theories.
For the linear combination of chiral-Lagrangian parameters entering the scattering length, this provides a first glimpse of how the reduced running of the coupling associated with the increase of $N_f$ affects the dynamics. In subsequent work, we will compute additional chiral-parameter combinations through both D-wave projections and further effective range parameters. Together, these will lead to a prediction for the two parameters of the electroweak chiral Lagrangian describing strongly coupled corrections to WW scattering.

\section{Electroweak Chiral Lagrangian and WW scattering}
The electroweak chiral Lagrangian (\EWC) allows for a systematic description of electroweak scale effects resulting from TeV scale physics \cite{Longhitano:1980iz,Appelquist:1993ka}.  This effective chiral Lagrangian must respect the $SU(2)_L \times U(1)_Y$ gauge symmetry.  
However, the chiral Lagrangian approximately respects a larger ``custodial" symmetry $SU(2)_L \times SU(2)_C$, which spontaneously breaks to the diagonal subgroup.  From the \EWC \cite{Longhitano:1980iz,Appelquist:1993ka}, the dominant contributions to WW scattering come from the terms
\beqs
\label{eq:Ltree}
\mathcal{L}^{W^4} &=&  - g^2 \tr[W_\mu, W_\nu]^2 \nonumber\\
&&+2ig \ \tr\Big((\partial_\mu W_\nu - \partial_\nu W_\mu)[W_\mu, W_\nu]\Big)\nonumber\\
&&+  \alpha_4 [\tr(V_\mu V_\nu)]^2 + \alpha_5 [\tr(V_\mu V^\mu)]^2,
\eeqs
where $V_\mu = (D_\mu U)U^\dag$,  $U(x)$ is the unitary matrix field that transforms under $SU(2)_L \times SU(2)_C$ (akin to the hadronic matrix of Goldstone fields), and the covariant derivative of $U(x)$ is given by
\beq
D_\mu U= \partial_\mu U + i g \frac{\vec{\tau}}{2}\cdot \vec{W}_\mu U - i g^\prime U \frac{\tau_3}{2} B_\mu.
\eeq
The $\alpha_4$ and $\alpha_5$ terms describe $\mathcal{O}(g^4)$ corrections to $WW$ scattering.

In Ref.~\cite{Eboli:2006wa}, Eboli et al found that with an integrated luminosity of $100 \; \text{fb}^{-1}$, and by considering both $W$ and $Z$ scattering, the LHC could place  $99\%$ CL bounds $-7.7 \times 10^{-3} < \alpha_4 < 15 \times 10^{-3}$ and $-12 \times 10^{-3} < \alpha_5 < 10 \times 10^{-3}$.
Because the fit to potential LHC data was made using only the terms of Eq. \eqref{eq:Ltree} at tree level, the $\alpha$-parameters are, in effect, defined at a low scale, incorporating all radiative corrections
including both standard-model corrections and new physics.  These parameters were also constrained by unitarity considerations in Refs.~\cite{Distler:2006if,Vecchi:2007na}. Other custodial-symmetry respecting coefficients in the \EWC have been constrained experimentally and do not lead to appreciable corrections to WW scattering \cite{Distler:2006if}.  More recent assessments of LHC constraints on vector boson scattering were performed in Ref.~\cite{Ballestrero:2011pe} and Ref.~\cite{Doroba:2012pd}. 

\begin{figure}[t]
  \includegraphics[width=85mm]{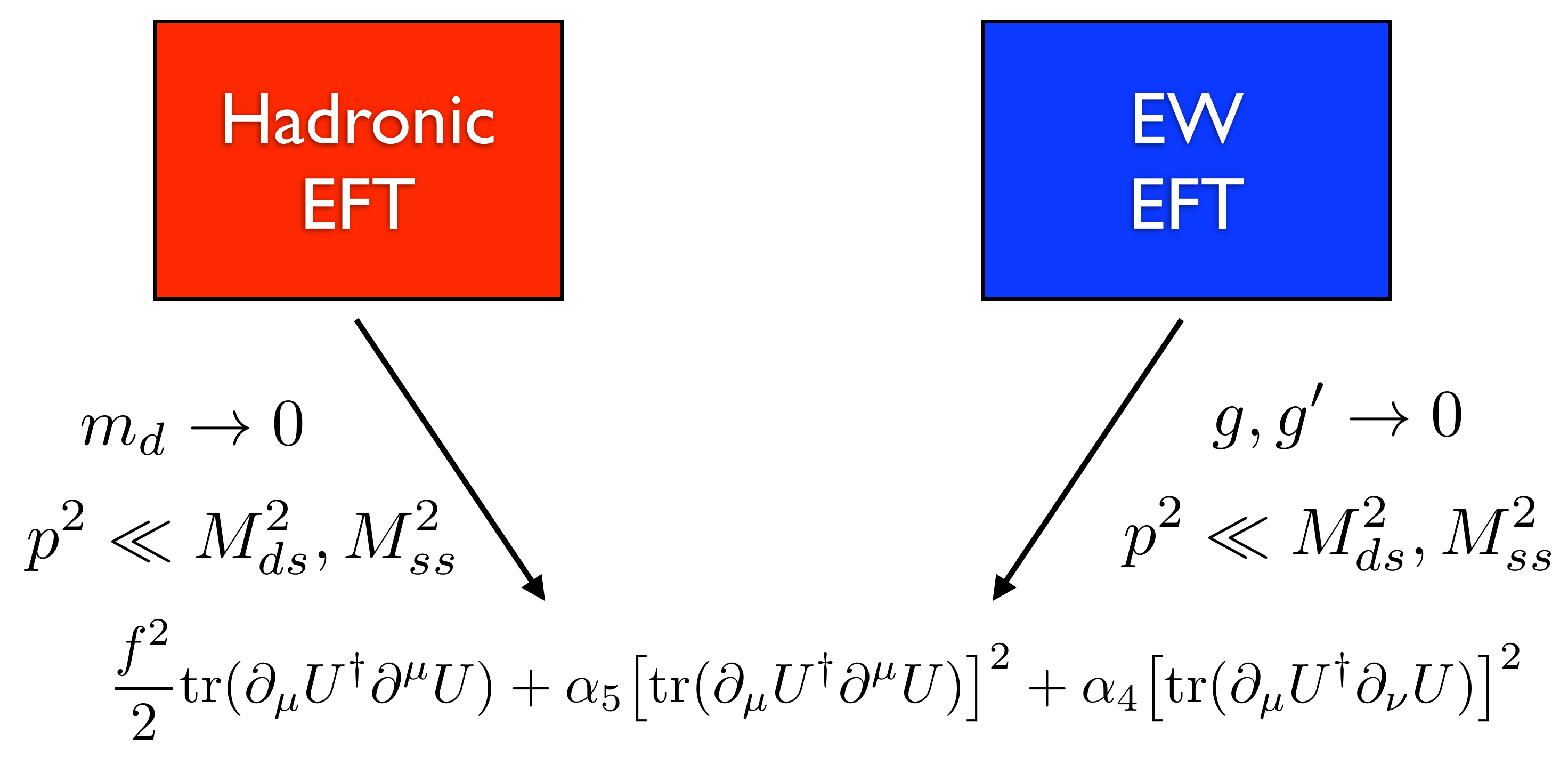}
  \caption{\label{fig:Had_EW}} Comparison between limits of the hadronic chiral Lagrangian with $N_f$ flavors \cite{Gasser:1983yg,Gasser:1984gg} and the electroweak chiral Lagrangian \cite{Longhitano:1980iz,Appelquist:1993ka}.  Symbolic quantities are defined in the text.
\end{figure}

In the limit $g, g^\prime \rightarrow 0$, the
\EWC reduces to the massless two-flavor hadronic chiral Lagrangian \cite{Gasser:1983yg}, as illustrated in Fig.~\ref{fig:Had_EW}, where
\beqs\label{eq:2_Flav_Rel}
\alpha_5 &=&  \frac{\ell_1}{4} + \mathcal{O}(g) \nonumber\\
 \alpha_4 &=& \frac{\ell_2}{4} + \mathcal{O}(g),
\eeqs
where $\ell_1$ and $\ell_2$ are the next-to-leading-order (NLO) Gasser-Leutwyler coefficients of the hadronic chiral Lagrangian with only derivative couplings,
\beq
\mathcal{L}_{NLO}^{\ell_1 \ell_2} = \frac{\ell_1}{4}[\tr(\partial_\mu U^\dag \partial^\mu U)]^2 + \frac{\ell_2}{4} [\tr(\partial_\mu U^\dag \partial_\nu U)]^2.
\eeq


To exhibit the flavor-dependent dynamics giving rise to $\alpha_4$ and $\alpha_5$, an extension to the multi-flavor hadronic chiral Lagrangian, with non-zero fermion masses is appropriate.
Theories with multiple flavors lead to additional Goldstone degrees of freedom, uneaten by the $W$ and $Z$ bosons, which, with a small mass, become PNGBs. Their contribution to physical phenomena are parametrized by the many low-energy constants (LECs) of the multi-flavor chiral Lagrangian. The presence of a finite fermion mass is also essential for the lattice simulations employed here.


For theories with $N_f \geq 4$ massive fermions, there are 9 LECs in the
NLO hadronic chiral Lagrangian, denoted by
$L_{0-8}$  \cite{Gasser:1984gg}. The LECs $L_{4-8}$ multiply terms
proportional to the fermion mass, while $L_{0-3}$ multiply terms that are
independent of the fermion mass. One way to relate these LECs to the \EWC is by
assigning electroweak quantum numbers to one fermion doublet among the
$N_f$ fermions, leaving the others neutral.

Although our lattice simulations are carried out with a common mass $m$ for all the fermions,
it is helpful for our discussion to temporarily assign a mass $m_d$ to the fermion doublet with electroweak quantum numbers and a mass $m_s \geq m_d$ to the remaining electroweak singlets. The former must be taken to zero, while the latter may or may not be taken to zero. The $N_f^2 -1$ PNGBs are then separated into the $3$ (with mass denoted by $M_{dd}$) which become massless when $m_d \rightarrow 0$ and are eaten by the gauge bosons, the $4(N_f - 2)$ composed of one electroweak-doublet fermion and one electroweak-singlet fermion (with mass denoted by $M_{ds}$), and the $(N_f - 2)^2$ composed of only electroweak-singlet fermions (with mass denoted by $M_{ss}$). All except the $3$ eaten modes can also get masses from standard-model and beyond-standard-model interactions not included here.


The relation of the 9 NLO LECs of the general-flavor hadronic Lagrangian to $\alpha_4$ and $\alpha_5$ in the electroweak Lagrangian is mapped out in Fig.~\ref{fig:Had_EW}. ``High-energy physics" including the PNGB particles composed of one or two of the electroweak-singlet fermions of mass $m_s$ are integrated out to describe observed low-energy physics on the order of the $W$ and $Z$ masses. Specifically, it is conventional to start with a set of $9$ LECs $L_i^r(\mu)$ defined at some scale $\mu$ small compared to the breakdown scale of the chiral Lagrangian. They are independent of PNGB masses, being defined in the zero-mass theory, and infrared finite because of the presence of the scale $\mu$.


Now suppose that the two-flavor Lagrangian of Eq. (4) refers to the two electroweak-doublet fermions. The two coefficients, $l^r_1$ and $l^r_2$, can also be defined at some scale $\mu$, and in the limit $M_{dd} = 0$. Since the PNGBs of mass $M_{ds}$ and $M_{ss}$ have been ``integrated out" (integrated into $l^r_1$ and $l^r_2$) these two coefficients do depend on \emph{these }masses. At NLO and with the PNGB masses small compared to the breakdown scale, they are related to the $L_i^r(\mu)$ by the addition of chiral logarithms arising from a loop of PNGBs composed of one EW-doublet fermion and one EW-singlet fermion \cite{Gasser:1984gg,Bijnens:2006zp}:
\beqs\label{eq:matching}
\ell_1^r(\mu,M_{ds})  &=& -2L_0^r(\mu) + 4L_1^r(\mu)+ 2L_3^r(\mu)  \nonumber\\
&&+ \frac{2-N_f}{24(32\pi^2)}\log \frac{M_{ds}^2}{\mu^2} \nonumber\\
\ell_2^r(\mu,M_{ds}) &=& 4L_0^r(\mu) + 4L_2^r(\mu) \nonumber\\
&&+ \frac{2-N_f}{12(32\pi^2)}\log \frac{M_{ds}^2}{\mu^2}.
\eeqs
 The coefficients of the the chiral logarithms are proportional to the expected counting factor of $4(N_f - 2)$.

 These quantities are related to the corresponding electroweak LECs $\alpha_5^r(\mu,M_{ds})$ and $\alpha_4^r(\mu,M_{ds})$ by Eq.~\eqref{eq:2_Flav_Rel}.
Then, a set of $\mu$-independent electroweak LECs $\widetilde{\alpha}_5(M_{dd},M_{ds})$ and $\widetilde{\alpha}_4(M_{dd},M_{ds})$ can be defined by adding chiral logarithms arising
at NLO from a loop of PNGBs composed of electroweak-doublet fermions of mass $m_d$:
\beqs\label{eq:Renorm_alpha}
\widetilde{\alpha}_5(M_{dd},M_{ds}) &=&  \frac{l_1^r(\mu,M_{ds})}{4} -   \frac{1}{384\pi^2}\log \frac{M_{dd}^2}{\mu^2}\nonumber\\
 \widetilde{\alpha}_4(M_{dd},M_{ds}) &=& \frac{l_2^r(\mu,M_{ds})}{4}  -   \frac{1}{192\pi^2}\log \frac{M_{dd}^2}{\mu^2},
\eeqs
where we have dropped $\mathcal{O}(g)$ corrections. These electroweak LECs have incorporated into them all strong-interaction effects including the PNGBs made purely of EW-doublet fermions, given at NLO by the explicit chiral-logarithm terms.

Comparison with the  $\alpha_4$ and $\alpha_5$ employed
by Eboli et al \cite{Eboli:2006wa} requires the inclusion of standard-model corrections not considered here.
Then the above $\log (M_{dd}^2$) dependence can be separated and incorporated into these
corrections (allowing the limit $M_{dd} \rightarrow 0$ to be taken there) since the resultant NGBs are
eaten by the $W$ and $Z$. We therefore subtract from Eq. \eqref{eq:Renorm_alpha}
the one-loop contributions arising from PNGBs of mass $M_{dd}$  together with a Higgs boson with reference mass $M_H$. From the first of Eq. \eqref{eq:Renorm_alpha}, we subtract  $(-1/384\pi^2)
[\log(M_{dd}^2/M_H^2)+O(1)]$, and from the second $(-1/192\pi^2)
[\log(M_{dd}^2/M_H^2)+O(1)]$, where the $O(1)$ constants are determined by the aforementioned one-loop calculation.
The remaining quantities are then
\beqs\label{eq:Subtractedalpha}
 \widetilde{\alpha}_5(M_{H},M_{ds}) &=&  \frac{l_1^r(\mu,M_{ds})}{4} -   \frac{\big[\log \frac{M_{H}^2}{\mu^2} +O(1)\big]}{384\pi^2}\nonumber\\
 \widetilde{\alpha}_4(M_{H},M_{ds}) &=& \frac{l_2^r(\mu,M_{ds})}{4}  -   \frac{\big[\log \frac{M_{H}^2}{\mu^2} + O(1)\big]}{192 \pi^2}.\nonumber\\
 \eeqs

 The subtracted LECs $\widetilde{\alpha}_4(M_{H},M_{ds})$ and $\widetilde{\alpha}_5(M_{H},M_{ds})$ are directly analogous to the conventionally defined $S$ parameter, as employed in Ref.~\cite{Appelquist:2010xv}. They can be determined experimentally and compared to lattice computations if the additional, standard-model corrections, with the same $M_H$, are also taken into account.
 For $N_f > 2$, they remain sensitive to the PNGB masses $M_{ds}$, and therefore the limit $m_{s} \rightarrow 0$ can be taken only if
additional physics provides a mass for these modes.

\section{Lattice Framework}
For the lattice calculation being performed here, all $N_f$ fermions are assigned a common mass $m$, and, as a result, have a common PNGB mass $M_P$. The goal is to compute the resultant $L_i^r(\mu)$ through pseudoscalar scattering, and then make use of Eqs. 5-7 with $M_P = M_{ds} = M_{dd}$, to determine the quantities $\widetilde{\alpha}_4(M_{H},M_{P})$ and $\widetilde{\alpha}_5(M_{H},M_{P})$ as a function of $M_P$. For $N_f = 2$, the limit $M_P \rightarrow 0$ is finite. For $N_f > 2$, the residual $\log M_P$ dependence means that the limit $m \rightarrow 0$ can be taken only if interactions not included here generate a mass for these modes.

\section{Maximal-isospin pseudoscalar scattering}
The study of pion scattering is well established in effective field theory, where the leading order (LO) result was first calculated by Weinberg \cite{Weinberg:1966kf} and the NLO chiral-perturbation-theory result was calculated by Gasser and Leutwyler \cite{Gasser:1984gg}.
Our focus here is on the scattering of two identical pseudoscalars within an $SU(2)_L \times SU(2)_R$ subgroup of the flavor symmetry, taken here to refer to the
fermion doublet assigned electroweak charges.  This process is often
referred to as ``I=2" scattering in the case of two flavors. Here, we refer to it as ``maximal-isospin"  scattering for $N_f$ fermion flavors with degenerate masses.

Low-energy pseudoscalar scattering can be parameterized in terms of the phase shift, $\delta$.  On the lattice, the accessible quantity is  $|\vec{k}| \cot \delta$, where $|\vec{k}|$ is the magnitude of the pseudoscalar 3-momenta.  The S-wave projection of this combination has a convenient effective-range expansion for small momenta given by
\beq\label{eq;Eff_Range_Expansion}
|\vec{k}| \cot \delta = \frac{1}{a_{PP}} + \frac{M_P^2r_{PP}}{2} \bigg(\frac{|\vec{k}|^2}{M_P^2}\bigg) + \mathcal{O}\bigg(\frac{|\vec{k}|^4}{M_P^4}\bigg),
\eeq
where $a_{PP}$ is the scattering length (in the particle physics convention), $r_{PP}$ is the effective range, and $M_P$ is the pseudoscalar mass. The expansion is valid for $|\vec{k}| \ll M_P$.

The NLO chiral expansion for the scattering length in the MI channel takes the form
\cite{Bijnens:2009qm, Bijnens:2011fm}
\begin{align}\label{eq:Pion_Scattering_Length_chi_scheme}
M_P a_{PP} = -\frac{M^2}{16\pi F^2}\Bigg\{1 + \frac{M^2}{16\pi^2 F^2}\Bigg[b_{PP}^r(\mu) && \nonumber\\ - ~\frac{2(N_f-1)}{N_f^2}
+ A(N_f) \log \Big(\frac{M^2}{\mu^2}\Big)\Bigg]\Bigg\},
\end{align}
where $M^2 = 2m\langle \overline{\psi}\psi\rangle / F^2$, with $F$  and $\langle \overline{\psi}\psi\rangle$ the pseudoscalar decay constant and chiral condensate at $m=0$, and where
\beqs
\label{eq:chi parameters}
A(N_f) &=& \frac{(2-N_f+2N_f^2+N_f^3)}{N_f^2}, \nonumber\\
b_{PP}^r(\mu) &=& -256\pi^2\big[(N_f-2)(L_4^r(\mu)-L_6^r(\mu))\nonumber\\
&&+L_0^r(\mu) + 2L_1^r(\mu)+2L_2^r(\mu)+L_3^r(\mu)\big].\nonumber\\
\eeqs
 Each $L^r_i(\mu)$ is defined at a scale $\mu$ in the ${\overline{MS}}$ scheme. The quantity
 $b_{PP}^r(\mu) + A(N_f) \log (M^2/\mu^2)$ is $\mu$-independent.

Similar chiral expansions of the pion mass and pion decay constant are given by \cite{Bijnens:2011fm}
\beqs
\label{eq:MandF}
M_P^2 &=& M^2\Bigg\{1+ \frac{M^2}{16\pi^2 F^2}\Bigg[b_M^r(\mu)+ \frac{1}{N_f}\log \frac{M^2}{\mu^2} \Bigg] \Bigg\} \nonumber\\
F_P &=& F\Bigg\{1+ \frac{M^2}{16\pi^2 F^2}\Bigg[b_F^r(\mu)- \frac{N_f}{2}\log \frac{M^2}{\mu^2} \Bigg] \Bigg\},\nonumber\\
\eeqs
where
\beqs
b_M^r(\mu) &=& 128\pi^2 [N_f(2L_6^r(\mu)-L_4^r(\mu)) \nonumber\\
&&+ 2L_8^r(\mu) - L_5^r(\mu)] \nonumber\\
b_F^r(\mu) &=& 64 \pi^2 [N_f L_4^r(\mu) + L_5^r(\mu)]\,.
\eeqs
The expressions for $M_P$ and $F_P$ are also $\mu$-independent.


Only the LECs $L_0$-$L_3$ contribute to WW scattering, (Eq.~\eqref{eq:matching}), since only they multiply operators that survive as $m \rightarrow 0$. But in general the scattering length computed here will give us the combination of parameters appearing in $b^r_{PP}$ (Eq.~\eqref{eq:chi parameters}), which include $L_4$ - $L_8$.  More data is required to separate these sets for general $N_f$.

For $N_f = 2$, however,
\beqs
b_{PP}^{r}(\mu) = -256\pi^2\big[L_0^r(\mu)
+ 2L_1^r(\mu)\nonumber\\ +2L_2^r(\mu)
+L_3^r(\mu)\big] \nonumber\\
= - 128 \pi^2\big[l_1^r(\mu)+ l_2^r(\mu)\big].
\eeqs
Both $l_1^r(\mu)$ and $l_2^r(\mu)$ are now $M_P$-independent since the only PNGBs are the three
composed of EW-doublet fermions, and therefore the chiral-logarithm terms in Eq. \eqref{eq:matching}
vanish. From Eq. \eqref{eq:Subtractedalpha}, we then have for the $N_f = 2$ case
\beqs
\label{eq:alphas2f}
\widetilde{\alpha}_4(M_H) + \widetilde{\alpha}_5(M_H) &=& -\frac{1}{512\pi^2}b_{PP}^{r}(\mu)
\nonumber\\
  &-&\frac{\big[\log \frac{M_{H}^2}{\mu^2} +O(1) \big]}{128\pi^2}\,.
\eeqs
Thus, for $N_f = 2$, a lattice calculation of $M_{P}a_{PP}$ as a function of $M$, with a fit using Eq. \eqref{eq:Pion_Scattering_Length_chi_scheme}, can directly determine $\widetilde{\alpha}_4(M_H) + \widetilde{\alpha}_5(M_H)$.

For $N_f = 2$, it is also worth noting that \emph{if} the analysis is taken to NNLO order in $M^2/F^2$, $\widetilde{\alpha}_4$ and $\widetilde{\alpha}_5$ can be determined separately, employing only the quantities $a_{PP}$, $M_P$ and $F_P$. The NLO LECs introduced above multiply single
chiral-logarithm terms at this order. While promising, the primary difficulty is having enough data within the chiral regime to constrain these terms accurately.

\section{Lattice details}
Lattice calculations are performed using $32^3 \times 64$ domain-wall lattices with the Iwasaki improved gauge action with a fifth dimensional length of $L_s = 16$ and a domain-wall height of $m_0 = 1.8$ \cite{Appelquist:2009ka,Appelquist:2010xv}.  By using domain-wall fermions, the calculation preserves exact flavor symmetry and chiral-breaking lattice spacing artifacts are $L_s$ suppressed ($m_{res} \ll m$).  This is performed for $N_f = 2$ at $\beta = 2.70$ and $N_f = 6$ at $\beta = 2.10$, where the beta values were chosen to match the IR scales of both theories and correspond to an inverse lattice spacing roughly five times the vector meson mass.  For both $N_f=2$ and $N_f = 6$, five mass points are analyzed with $m =m_f + m_{res}$, where $m_f = 0.010, 0.015, 0.020, 0.025,0.030$.

\section{Finite Volume Method}
In Euclidean space, scattering phase shifts can be extracted on the lattice only by calculating the total energy of two hadrons in a finite volume \cite{Luscher:1986pf}.  In practice, this is accomplished by calculating the four-point correlation function and exploring the long time behavior
\begin{align}\label{eq:Four_Point}
\tr(\pi(t)^2)-\tr(\pi(t))^2 \sim e^{-E_{PP} t},
\end{align}
where the zero-momentum projected operator $\pi(t)$ is related to the pseudoscalar two-point function by $C_P(t)=\tr(\pi(t))$, and $E_{PP}^2 = 4(|\vec{k}|^2 + M_P^2) $, with $|\vec{k}|$ being the center-of-mass scattering momentum.   In this work, we restrict ourselves to S-wave scattering by projecting each pseudoscalar correlator onto zero momentum.  For a finite volume, only discretized values of the scattering interaction are allowed.   This scattering momentum is related to the phase shift by \cite{Luscher:1986pf}
\beq \label{eq:Luscher_Rel}
|\vec{k}| \cot \delta = \frac{1}{\pi L}S\Bigg(\frac{|\vec{k}|^2L^2}{4\pi^2}\Bigg),
\eeq
where the function $S(\eta)$ is given by the regularized zeta function \cite{Beane:2003da}
\beq\label{eq:S}
S(\eta) = \sum_{\mathbf{j} \neq 0}^{\Lambda} \frac{1}{|\mathbf{j}|^2 - \eta}-4\pi\Lambda.
\eeq

\section{Calculation and fitting}
The calculation yields two quantities: the four-point pseudoscalar correlation function, $C_{PP}(t)$, and the two-point correlation function $C_P(t)$.  The $F_P$ values used in this work were previously calculated in Ref.~\cite{Appelquist:2009ka}.  The long time behavior of the two correlators differ due to the extra backward propagating pseudoscalar that exists in $C_{PP}$(t) and as a result, the long time behavior of the midpoint-symmetric correlators is given by
\beqs
C_{PP}(t) &\rightarrow& A + B\cosh(E_{PP}t) \nonumber\\
C_{P}(t) &\rightarrow& C\cosh(M_{P}t)\,.
\eeqs
For the pseudoscalar mass, a hyperbolic cosine fit was performed, where
\beq \label{eq:cosh}
\cosh(M_{P}) = \frac{C_P(t+1)+C_P(t-1)}{2C_P(t)}.
\eeq
A different fitting form was used for $E_{PP}$, namely
\beq\label{eq:apcosh}
\cosh(E_{PP}) = \frac{C_{PP}(t+2)-C_{PP}(t-2)}{2(C_{PP}(t+1)-C_{PP}(t-1))}.
\eeq
Constant fits were performed to the inverse hyperbolic cosine of Eq.~\eqref{eq:cosh} and Eq.~\eqref{eq:apcosh}  over a window of roughly 40 time slices in the plateau region.  In addition to statistical errors, systematic effects due to placement of the fitting window were also examined by varying the window by $\pm2$ time slices on each side.
\begin{figure}[t]
\includegraphics[width=85mm]{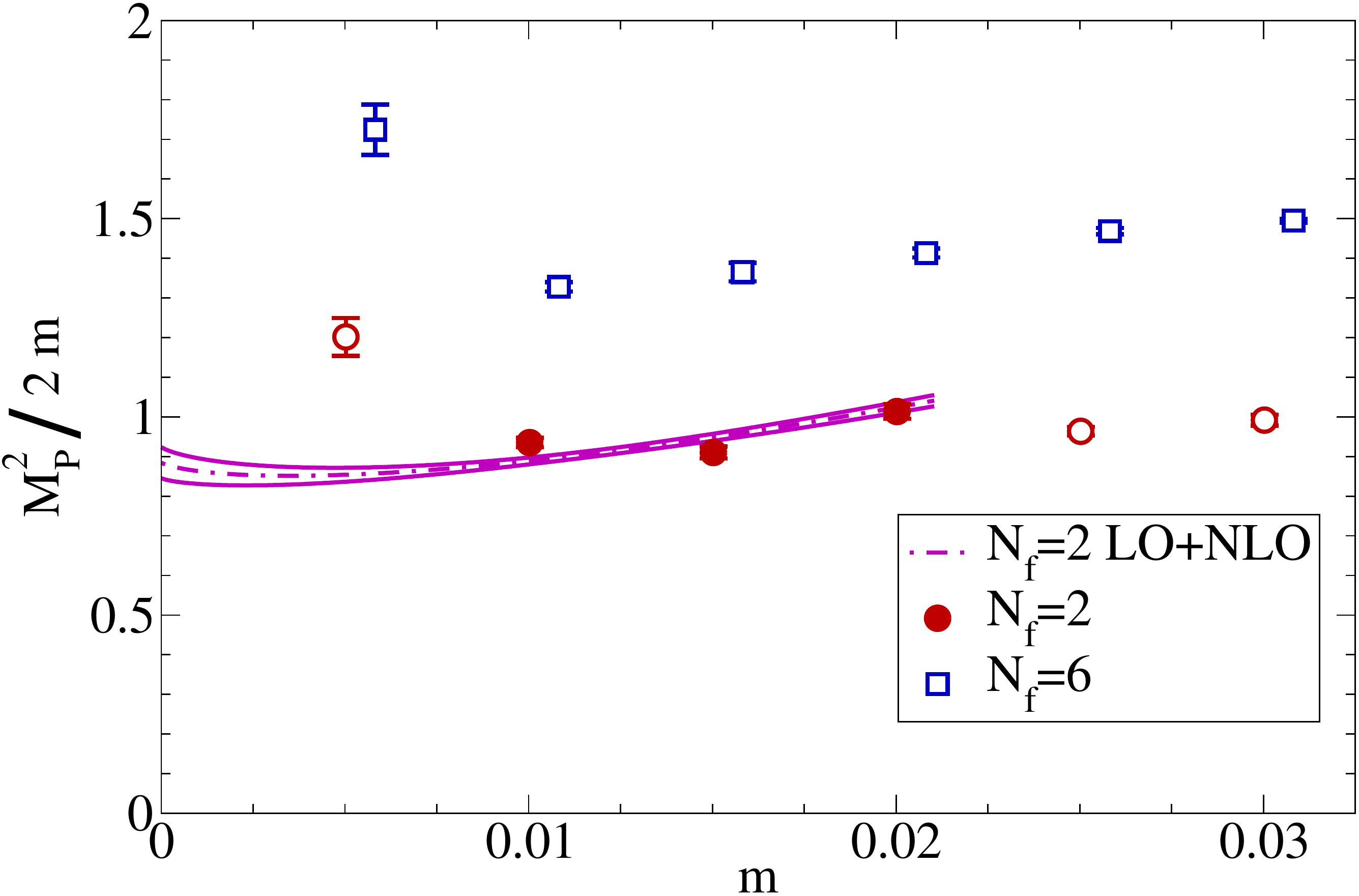}
\includegraphics[width=85mm]{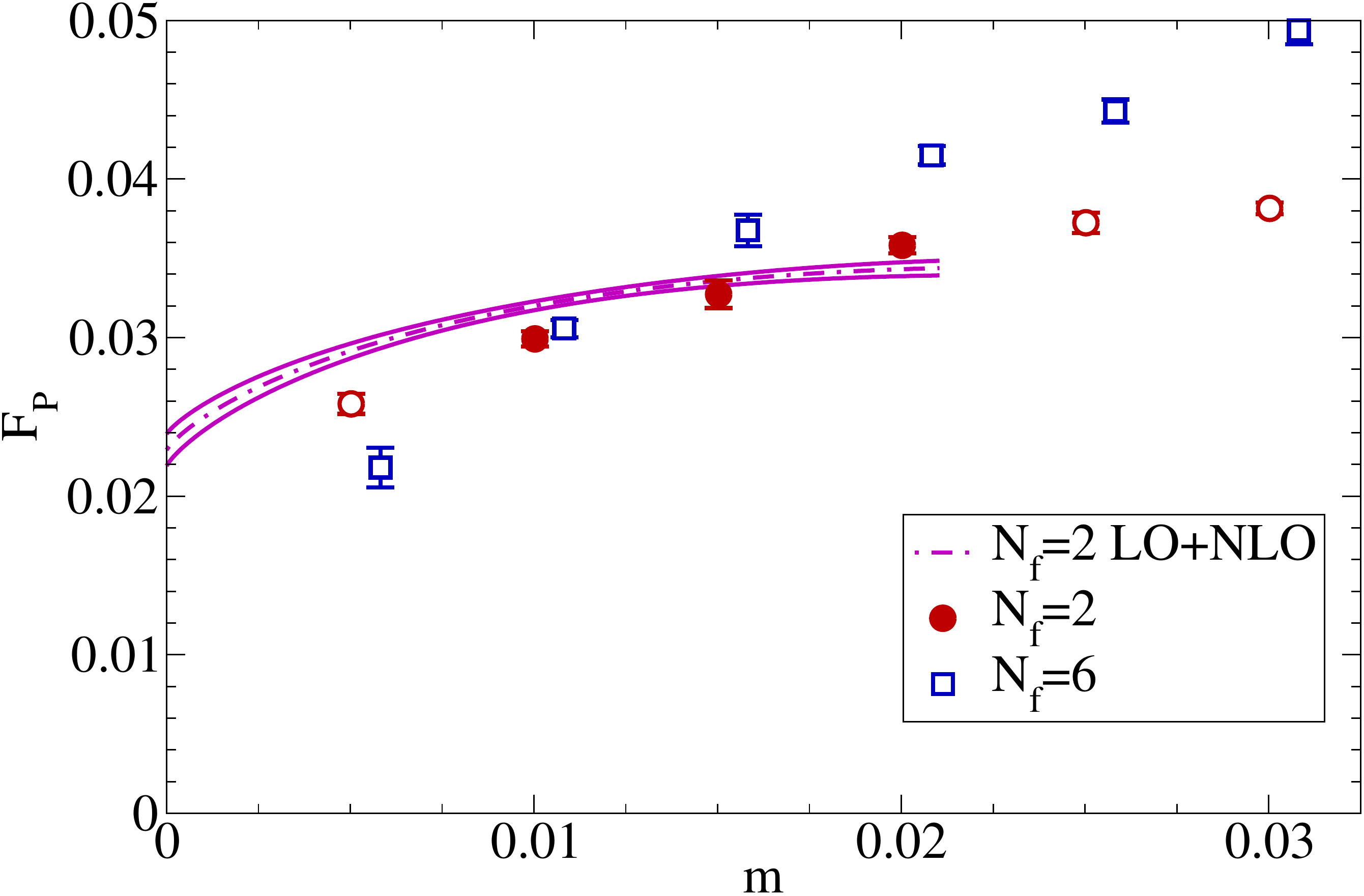}
  \includegraphics[width=85mm]{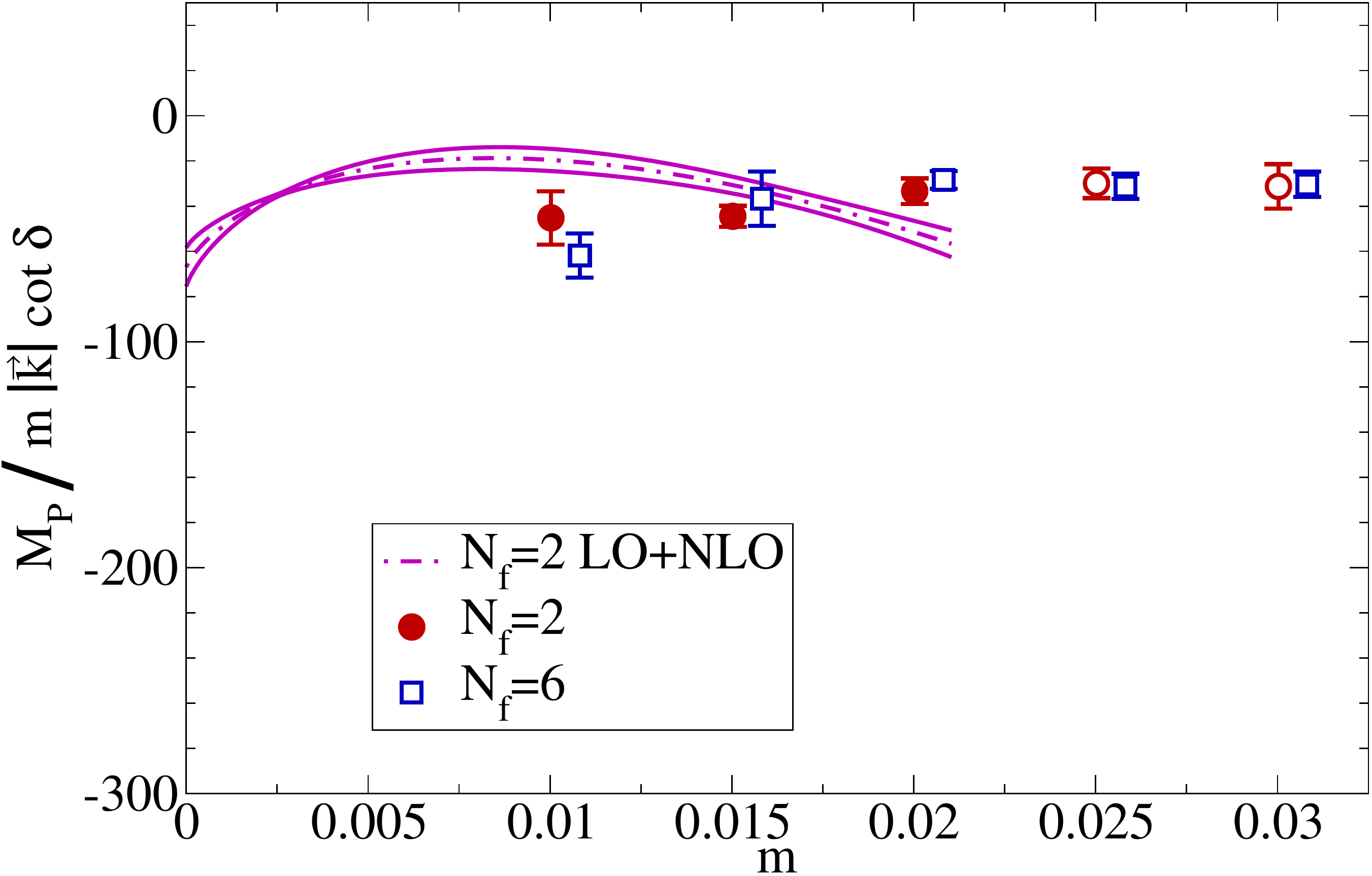}
  \caption{\label{fig:I2_mf} Plots  of $M_P^2/2m$, $F_P$ and $M_P/m|\vec{k}|\cot\delta \simeq M_{P}a_{PP}/m$ versus $m$ (in lattice units). The error bars are statistical plus systematic. The red circles represent the $N_f =2$ data and the blue squares represent the $N_f =6$ data. The fits for $N_f = 2$ are made using only the solid red points.}
\end{figure}

\section{Scattering results}
In our lattice calculations, $|\vec{k}| \cot \delta$ is extracted from Eq.~\eqref{eq:Luscher_Rel}.   For
Eq.~\eqref{eq:Pion_Scattering_Length_chi_scheme} to be applicable, the effective range expansion, Eq.~\eqref{eq;Eff_Range_Expansion}, is needed and $|\vec{k}|^2/M_P^2$ must be sufficiently small. The data suggest that the approximation $M_P/|\vec{k}| \cot \delta \approx M_P a_{PP}$ is valid within 9\%, and thus Eq.~\eqref{eq:Pion_Scattering_Length_chi_scheme} should be applicable to the lattice results. Our results for pseudoscalar scattering in the MI channel are given as a function of $m$ in  the bottom graph of Fig.~\ref{fig:I2_mf}. We first analyze the $N_f = 2$ results and then discuss the difference between
 $N_f = 2$ and $N_f = 6$.


 \subsection{$\bf{N_f = 2}$}

 To determine $b_{PP}^r(\mu)$ for $N_f = 2$, we carry out a combined fit using NLO chiral perturbation
 to $M_{P}a_{PP}$ (Eq. \eqref{eq:Pion_Scattering_Length_chi_scheme}), $M_P^2$ (Eq. \eqref{eq:MandF}), $F_P$ (Eq. \eqref{eq:MandF}), and the chiral condensate. The latter three quantities were computed and analyzed
 in Ref.~\cite{Appelquist:2009ka}, where it was concluded that for the range of $m$ values employed there, NLO chiral perturbation theory provides an acceptable fit allowing a reliable extrapolation to
 $m=0$, determing the extrapolated decay constant $F$ and chiral condensate $\langle \overline{\psi}\psi\rangle$.

\begingroup
\squeezetable
\begin{table}[h]
\caption{\label{tab:Nf=2_chiral_fit} Results (in lattice units) of a combined NLO fit for $N_f = 2$ to $M_P a_{PP}$, $M_P^2$, $F_P$, and the chiral condensate.}
\begin{ruledtabular}
\begin{tabular}{l||c|c|c||c}
& $m_f$=0.01--0.015 & \textbf{$m_f$=0.01--0.02} & $m_f$=0.015--0.02 &
  Ref.~\cite{Appelquist:2009ka} \\
\hline
$b_C^r(F)$    & 80(10)     & \textbf{110.7(9.0)} & 198(60)    &
  91(29)   \\
$b_F^r(F)$    & 5.20(28)   & \textbf{5.38(15)}   & 5.09(58)   &
  5.70(27) \\
$b_M^r(F)$    & -2.36(21)  & \textbf{-1.74(24)}  & 0.5(2.2)   &
  -2.22(62) \\
$b_{PP}^r(F)$ & -17.32(88) & \textbf{-16.89(59)} & -13.3(3.8) &
  --- \\
$F$           & 0.0220(16) & \textbf{0.0229(10)} & 0.0262(35) &
  0.0209(41) \\
$\pbp/F^2$    & 1.049(74)  & \textbf{0.885(39)}  & 0.65(10)   &
  0.99(17) \\
\hline
$\chi^2 / \mathrm{dof}$ & 16 / 2 & \textbf{83 / 6} & 13 / 2 &
  26 / 4
\end{tabular}
\end{ruledtabular}
\end{table}
\endgroup

 For the present fit, we choose $\mu =F \simeq 250$ GeV (In Ref.~\cite{Appelquist:2009ka}, the scale $\mu = 4 \pi F$ was used.) The (combined) fits for $M_P^2/2m$, $F_P$ and
 $M_P/m|\vec{k}| \cot \delta \simeq M_P a_{PP}/m$ are shown in Fig.~\ref{fig:I2_mf}. Only the points $m_f = 0.01 - 0.02$ (shown in solid red) are used in the fit. For all quantities, NLO chiral perturbation theory again provides an acceptable fit as in Ref.~\cite{Appelquist:2009ka}.

 The fit parameters are shown in bold face in the central column of
 Table~\ref{tab:Nf=2_chiral_fit}. In addition, fit parameters for two other $m_f$ ranges are shown,
 and used to estimate systematic errors. From the table, the quantity of interest here is
\beqs
\label{eq:l_values}
b_{PP}^{r}(\mu = F) &=& - 128 \pi^2 \big[l_1^{r}(\mu = F) + l_2^{r}(\mu= F)\big] \nonumber\\
\nonumber\\
 &=& - 16.89 \pm 0.59_{-0.43}^{+3.59} \quad \frac{\chi^2}{\text{dof}}=\frac{83}{6},\nonumber\\
\eeqs
where we have used the central value from the fit range $m_f = 0.01 - 0.02$. The errors are statistical plus systematic. This result is consistent with previous lattice simulations of I=2 pion-pion scattering \cite{Yamazaki:2004qb,Beane:2005rj,Beane:2007xs,Feng:2009ij,Dudek:2010ew,Beane:2011sc,Yagi:2011jn}, other 2+1 lattice QCD determinations of these LECs \cite{Aoki:2009qn,Beane:2011sc}, and QCD phenomenology.  Lighter mass ensembles will be required for future calculations to achieve higher precision.

From Eq. \eqref{eq:alphas2f}, we then have, for $N_f = 2$,
\beqs
\label{eq: alpha values}
\widetilde{\alpha}_4(M_H) + \widetilde{\alpha}_5(M_H) &=& (3.34 \pm  0.17_{-0.71}^{+0.08}) \times 10^{-3} \nonumber\\
&-&\frac{\big[ \log \frac{M_H^2}{F^2}+O(1)\big]}{128 \pi^2} \,,
\eeqs
where the errors are again statistical plus systematic.
Recall that the expression in the second line arises from having removed the one-loop contribution arising from the eaten Goldstone bosons and a Higgs boson with reference mass $M_H$. For a wide range of $M_H$, it is  $O(10^{-3})$ or smaller.

Once the additional one-loop standard-model corrections with the same $M_H$ are included, the above result can be compared directly with LHC data. If the standard-model corrections are also of order
$10^{-3}$ or smaller, then the full set of $O(g^4)$ contributions will fall comfortably
within the Eboli et al bounds based on $100$ fb$^{-1}$ of LHC data. Future measurements will have to be
more precise to compare meaningfully with Eq. \eqref{eq: alpha values}.

                                    \subsection{$\bf{N_f = 6}$}

The $N_f = 6$ data for $M_P/m|\vec{k}| \cot \delta \simeq M_P a_{PP}/m$ displayed in Fig.~\ref{fig:I2_mf}
lies statistically on top of the $N_f = 2$ data. A NLO fit using Eq. \eqref{eq:Pion_Scattering_Length_chi_scheme}
would nevertheless lead to a different value for $b^r_{PP}(\mu)$ because the chiral-logarithm term in this expression has a much larger coefficient, growing linearly with $N_f$. However, the larger coefficient indicates that it is very unlikely that Eq. \eqref{eq:Pion_Scattering_Length_chi_scheme} can be employed in the $N_f = 6$ case for the existing range of $m$ values. This point was already made in Ref.~\cite{Appelquist:2009ka} where $F_P$, $M_P$ and the chiral condensate were computed.
The NLO expression for $F_P$ (Eq. \eqref{eq:MandF}) also has a chiral-logarithm term with a coefficient growing linearly with $N_f$.

Nevertheless, our results for $N_f = 6$ do indicate an interesting trend, which can be seen more easily by
plotting the data in a different way. If the NLO chiral expansion Eq. \eqref{eq:Pion_Scattering_Length_chi_scheme} is reorganized in terms of the
physical values $M_P$ and $F_P$, it takes the form
\begin{align}\label{eq:Pion_Scattering_Length}
M_P a_{PP} =& -\frac{M_P^2}{16\pi F_P^2}\Bigg\{1 + \frac{M_P^2}{16\pi^2 F_P^2}\Bigg[b_{PP}^{r\prime}(\mu) \nonumber\\
&-\frac{2(N_f-1)}{N_f^2} + A^{\prime}(N_f) \log \Big(\frac{M_P^2}{\mu^2}\Big)\Bigg]\Bigg\}\,,
\end{align}
where
\beqs \label{eq:l_pi_pi}
A^{\prime}(N_f) &=& \frac{2(1-N_f+N_f^2)}{N_f^2} \nonumber\\
b_{PP}^{\prime r}(\mu) &=& -256\pi^2\big[L_0^r(\mu) + 2L_1^r(\mu)+2L_2^r(\mu)\nonumber\\
&&+L_3^r(\mu)-2L_4^r(\mu)-L_5^r(\mu)+2L_6^r(\mu)\nonumber\\
&&+L_8^r(\mu)\big]\,.
\eeqs
With $M_P a_{PP}$ expressed in this way, the coefficient of the logarithmic term remains finite as $N_F$ increases, and the LEC combination $b_{PP}^{\prime r}(\mu)$ contains no explicit $N_f$ dependence.

\begin{figure}[t]
   \includegraphics[width=85mm]{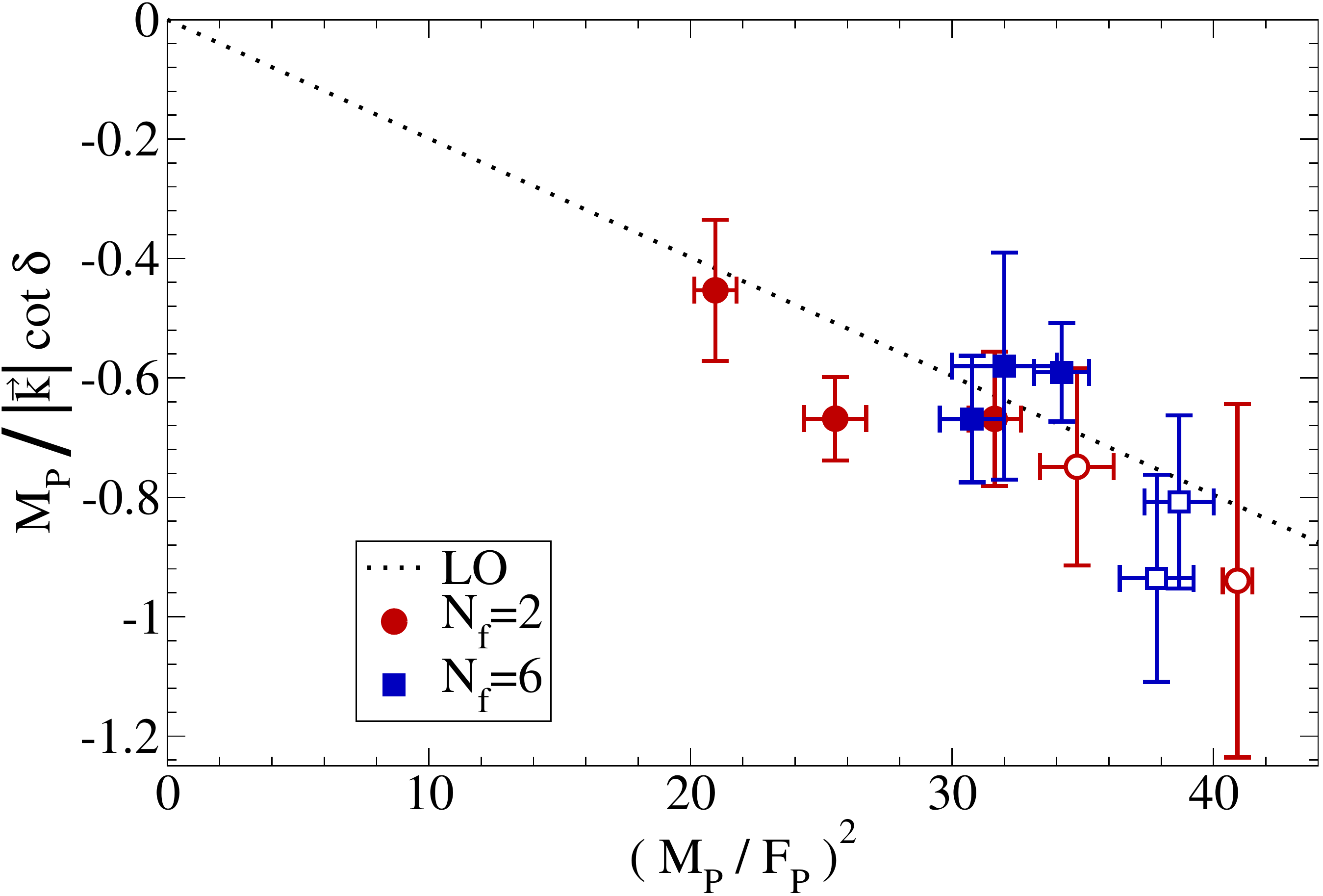}
  \caption{\label{fig:I2} Plot of $M_P/|\vec{k}|\cot\delta \simeq M_{P}a_{PP}$ vs. $(M_P/F_P)^2$. The error bars are statistical plus systematic.  The red circles represent the two-flavor data and the blue squares represent the six-flavor data.  The dashed line is the LO \CPT result (zero parameter fit). Larger negative results correspond to more repulsive scattering.}
\end{figure}

The above expansion, while not a priori more reliable for $N_f = 6$ than Eq. \eqref{eq:Pion_Scattering_Length_chi_scheme}, suggests that a plot of $M_P a_{PP} \simeq M_P/|\vec{k}| \cot \delta $  versus the physical quantity $M_P^2/F_P^2$ could be revealing. This is done in Fig.~\ref{fig:I2}, where the solid-color points correspond to the range $m_f = 0.01 - 0.02$. A small upward shift of the $N_f = 6$ points relative to $N_f = 2$ is indicated. Since a negative value for
$a_{PP}$ corresponds to a repulsive interaction, the data indicate that the $N_f = 6$ theory is somewhat less repulsive than the $N_f =2$ theory for pseudoscalar scattering in the maximal-isospin channel.

The dashed line, representing the LO expression $-M_P^2 / 16 \pi^{2} F_P^2$, is a reasonably good first approximation to the data for both $N_f=2$ and $N_f=6$. For $N_f = 2$, the data show that the effect of the NLO term is to make the interaction more repulsive. The quantity in square brackets in Eq. \eqref{eq:Pion_Scattering_Length} is positive and of order unity within the range shown.  A fit to just $M_{P}a_{PP}$ with $\mu = F$ leads to the value $b_{PP}^{\prime r}(\mu =  F) = -4.67\pm 0.65^{+1.06}_{-0.05}$.  
Clearly there is some cancelation between this term and the chiral logarithm.  Nonetheless, this $b_{PP}^{\prime r}$ value (when combined with the $b_{M}^{r}$ and $b_F^r$ values in Table~\ref{tab:Nf=2_chiral_fit}) is consistent with the $b_{PP}^{r}$ value in Eq.~\eqref{eq:l_values}.

 For $N_f = 6$, the data is even closer to the LO dashed line, suggesting that NLO perturbation theory in the form of Eq. \ref {eq:Pion_Scattering_Length} might again be reliable. If this expression is used to fit the $N_f = 6$ data, then the quantity
in square brackets is again positive and of order unity within the range shown, but somewhat smaller in magnitude than for $N_f=2$.  Since we don't yet know the precise value of $F$ in lattice units for 
$N_f = 6$,
we carry out the NLO fit using  the scale $\mu = 0.023 a^{-1}$ ($F$ for 
$N_f=2$).  The fit  leads to  $b_{PP}^{\prime r}(\mu= 0.023 a^{-1} \simeq F) = -7.81\pm 0.46^{+1.23}_{-0.56}$, larger in magnitude than for $N_f =2$. There is now more cancelation between this term and the chiral logarithm
than for $N_f = 2$.

\begin{figure}[t]
   \includegraphics[width=85mm]{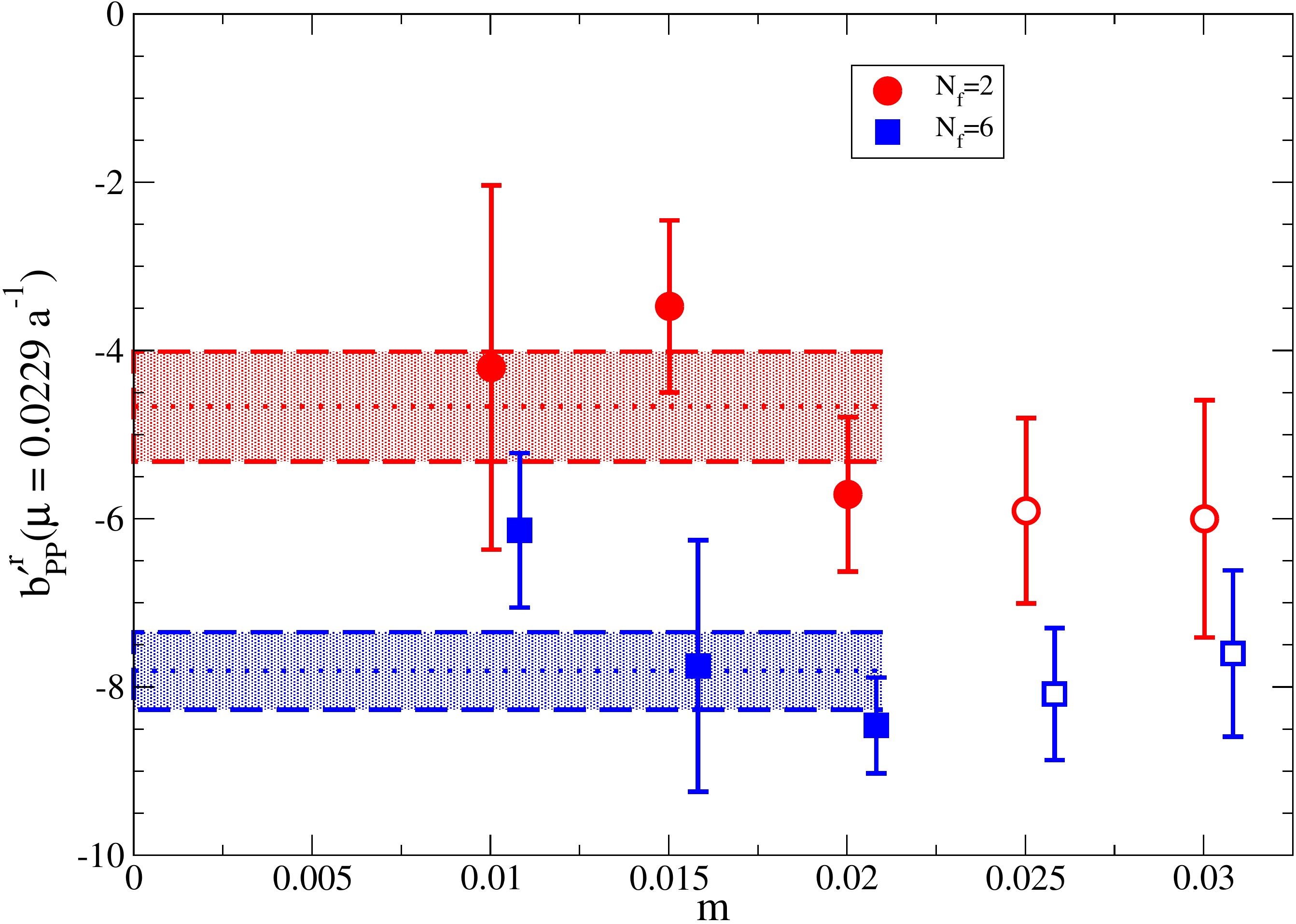}
  \caption{\label{fig:bPP} Chiral parameter $b_{PP}^{\prime r}$ versus fermion mass $m$ for $N_f = 2$ and $N_f = 6$. }
\end{figure}

The above values of $b_{PP}^{\prime r}$ emerge from a fit of Eq. \eqref{eq:Pion_Scattering_Length} to each of the three lightest data points (corresponding to $m_f = 0.01 - 0.02$), with a fixed choice $\mu = 0.023a^{-1} \simeq F$. A plot of the resultant value of $b_{PP}^{\prime r}$ versus $m$ (Fig. \ref{fig:bPP}), shows that $b_{PP}^{\prime r}(\mu= 0.023 a^{-1}\simeq F)$ is relatively independent of $m$ for both $N_f = 2$ and $N_f = 6$ as expected if NLO perturbation theory is reliable. The evident shift going from $N_f = 2$ to $N_f = 6$ is interesting since this quantity is contains LEC's that enter into WW scattering through Eq. \eqref{eq:l_pi_pi}.

 It is not yet clear whether this fit can be trusted for $N_f = 6$, but even if it can, the resultant value for $b_{PP}^{r\prime} (\mu= 0.023 a^{-1}\simeq F)$ determines only the combination of LECs in Eq. \eqref{eq:l_pi_pi}, which includes $L_i^r(\mu)$ values not directly relevant to WW scattering. Further calculations will be necessary to isolate $\widetilde{\alpha}_4(M_{H},M_P = M_{ds})$ and  $\widetilde{\alpha}_5(M_{H}, M_P = M_{ds})$ (Eq. \eqref{eq:Subtractedalpha}). This will then describe the effect of beyond-standard-model physics for a
range of PNGB masses $M_P$.

\section{Summary and Discussion}

Using lattice simulations, we have computed pseudoscalar-pseudoscalar scattering in the maximal isospin channel for an $SU(3)$ gauge theory with two and six fermion flavors in the fundamental representation. Our calculation of the S-wave scattering length was then related to the next-to-leading order (NLO) corrections to WW scattering through the low-energy coefficients of the chiral Lagrangian. For $N_f = 2$, our result for the scattering length agreed with previous calculations, showing an increase in repulsion due to the NLO corrections. For WW scattering, we obtained an estimate for $\widetilde{\alpha}_4(M_H) + \widetilde{\alpha}_5(M_H)$ (Eq. \eqref{eq: alpha values}) describing deviations from the standard model.

Six-flavor scattering showed a somewhat less repulsive NLO interaction than its two-flavor counterpart for a fixed ratio of the pseudoscalar mass to its decay constant. The range of fermion masses employed so far does not allow a clearly reliable use of chiral perturbation theory. Also, the appearance of more terms in the hadronic chiral lagrangian for six flavors does not allow the extraction of only the combination of parameters entering WW scattering. Further simulations of additional low-energy scattering parameters at lower fermion-mass values  will be required to complete this study.

\bigskip
\paragraph{\textbf{Acknowledgements}} This work was performed with the aid of Chroma \cite{Edwards:2004sx} and CPS.  We thank Tom Luu, Andr\'e Walker-Loud, and Brian Tiburzi for helpful insight throughout this work.  We thank the LLNL Multiprogrammatic and Institutional Computing program for time on the BlueGene/L supercomputer and on the Hera, Atlas, and Zeus computing clusters along with funding from LDRD 10-ERD-033.  This work was supported by the NNSA and Office of Science of the U.S. Department of
 Energy, and by the U.S. National Science Foundation.  Fermilab is operated by Fermi Research Alliance, LLC, under Contract DE-AC02-07CH11359 with the United States Department of Energy.
 
Preprint Numbers:

LLNL-JRNL-499587

FERMILAB-PUB-12-012-T

\bibliography{LSD-Pi_Pi}

\end{document}